\RequirePackage{fix-cm}
\documentclass[twocolumn]{svjour3}          
\smartqed  
\usepackage{graphicx}
\usepackage{hhline}
\usepackage{amssymb}
\usepackage{url}

\begin{document}

\title{Laboratory investigations on the resonant feature of `dead water' phenomenon}

\titlerunning{Laboratory investigations of `dead water'}        

\author{Karim Medjdoub \and Imre M. J\'anosi \and Mikl\'os Vincze}


\institute{K. Medjdoub \at
              von K\'arm\'an Laboratory for Environmental Flows, P\'azm\'any P. stny. 1/a, H-1117 Budapest, Hungary \\
              \email{karim26@caesar.elte.hu}           
           \and
M. Vincze \at
              MTA-ELTE Theoretical Physics Research Group, P\'azm\'any P. stny. 1/a, H-1117 Budapest, Hungary \\
              von K\'arm\'an Laboratory for Environmental Flows, P\'azm\'any P. stny. 1/a, H-1117 Budapest, Hungary \\
              \email{mvincze@general.elte.hu}   
                         \and
I. M. J\'anosi \at
            Department of Physics of Complex Systems, ELTE Eötvös Loránd University, Pázmány sétány 1/a, H-1117, Budapest, Hungary \\
            Max Planck Institute for the Physics of Complex Systems, N\"othnitzer Str. 38, D-01187 Dresden, Germany \\
            \email{imre.janosi@ttk.elte.hu}   
}

\date{Received: date / Accepted: date}

\maketitle

\begin{abstract}
Interfacial internal wave excitation in the wake of towed ships is studied experimentally in a quasi-two layer fluid. At a critical `resonant' towing velocity, whose value depends on the structure of the vertical density profile, the amplitude of the internal wave train following the ship reaches a maximum, in unison with the development of a drag force acting on the vessel, known in the maritime literature as `dead water'. The amplitudes and wavelengths of the emerging internal waves are evaluated for various ship speeds, ship lengths and stratification profiles. The results are compared to linear two- and three-layer theories of freely propagating waves and lee waves. 
We find that despite the fact that the observed internal waves can 
have considerable amplitudes, linear theories can still provide a surprisingly adequate description of subcritical-to-supercritical transition and the associated amplification of internal waves.
\end{abstract}

\section{Introduction}
\label{intro}
When a ship is travelling through a strongly stratified water body a certain amount of its kinetic energy is being used up for the excitation of internal waves in its wake, hardly noticeable from the surface, yet perceived as a drag force acting on the vessel.
For centuries, the phenomenon has been known to Norwegian seamen as ``d\"odvand'' or dead water. In the fjords of the Scandinavian coastline fresh water from slow glacier runoff gently sets on top of the saline seawater without substantial mixing and hence nearly jump-wise vertical density profiles can develop \cite{glacier}. These circumstances facilitate particularly strong dead water effect associated with large amplitude wave activity along the internal density interface (pycnocline). 

In the logbook of the 1893-96 Norwegian Polar Expedition Arctic explorer Fridtj\"of Nansen reported experiencing marked dead water drag on board research vessel \emph{Fram} that reduced the ship's speed to a fifth part. Nansen's original observations were further analyzed by (the then-PhD student) V. W. Ekman, who, in order to understand the phenomenon, has conducted laboratory experiments in a quasi-two-dimensional wavetank, filled up with a two-layer working fluid consisting of saline- and freshwater.  In these measurements a scale model of \emph{Fram} was towed along the surface subjected to either constant or gradually changing force as control parameter, against which the model's velocity $U$ was measured \cite{ekman}.
Ekman found that for smaller towing forces (and smaller $U$) dead water drag $F_{\rm dw}$ follows a quadratic scaling $F_{\rm dw} \approx \zeta U^2$ up to a certain threshold, where coefficient $\zeta$ is a function of the density profile $\rho(z)$ and the wetted area of the ship.

\begin{figure}
\centering
\noindent\includegraphics[width=0.6\columnwidth]{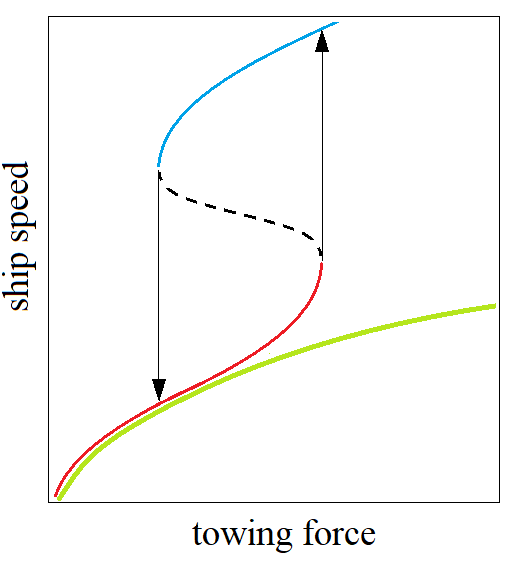}
\caption{Hysteresis of ship speed -- as observed and discussed by Ekman in his original work -- illustrated graphically as a function of changing towing force (decreasing branch: blue, increasing branch: red). The unstable branch, where the dead water phenomenon occurs, is marked with a dashed line. The green curve represents the (inverse of) subcritical relationship $F_{\rm dw} \approx \zeta U^2$.}
\label{hyst}
\end{figure}

However, when $U$ reaches a critical value of $U\approx 0.8 \cdot c_0^{(2)}$, where $c_0^{(2)}$ denotes the long-wave velocity of interfacial waves on the pycnocline (to be discussed later) the corresponding $F_{\rm dw}$ starts to decrease significantly. Then, in Ekman's ``changing force''-type experiments a hysteresis was encountered, as sketched in Fig. \ref{hyst}. When the gradually increasing towing force reached a certain tipping point (the right end of the blue branch in Fig. \ref{hyst}) the speed of the ship suddenly jumped to a higher value. An analogous drop of $U$ could be observed when decreasing the towing force in time (red branch of Fig. \ref{hyst}). The unstable branch (dashed line) is inaccessible to the system for any prescribed towing force. When initiating experiments with a \emph{constant} force within this hysteretic regime, Ekman found that the ship's velocity $U$ exhibited large fluctuations, comparable to the mean value \cite{ekman}.

If one intends to explore the dynamics in this unstable branch it is therefore beneficial to conduct experiments in which, instead of the applied force, the towing velocity $U$ is prescribed. Such settings are common in the laboratory modeling of lee-wave dynamics (see, e.g. \cite{leewave_exp,knigge,vosper1999}), where obstacles of various shapes are being towed at the surface (or at the bottom) in a tank of stratified working fluid and the properties of the generated internal waves, hydraulic jumps and `wave rotors' are evaluated \cite{vosper2004,sachs1,sachs2}.

In the present experimental work we follow a similar approach to address the dead water phenomenon, by pulling a ship model at a constant speed and -- instead of the experienced drag -- analyzing the properties of the excited interfacial waves on the pycnocline. Here the `critical regime' is characterized by wave patterns whose vertical extent is comparable to the height of the upper layer. The fact that wave amplitudes show strong dependence on the wave velocity (that is set by ship speed $U$) implies that the waves are of nonlinear nature \cite{yuan}. Even for velocities where no wave trains are observed, the localized bump following the ship at the pycnocline resembles the solitary wave solutions of the nonlinear Korteweg--de Vries equation \cite{kdv,boschan}.

A large pool of theoretical, numerical and experimental studies exists discussing the properties of internal waves emerging in the dead water problem addressing nonlinear wave excitation both in the subcritical \cite{grue_sub} and supercritical \cite{grue_super} regimes, as well as the applicability of the theoretical findings to observational data, including the historical logs of \emph{Fram} \cite{grue_fram}. 

Despite of the obvious nonlinearity of the problem, as far as the interfacial wave forms are concerned, the effect of nonlinear corrections is found to be surprisingly minor (see, e.g. Fig. 5 of \cite{grue_super}). In a series of earlier laboratory experiments on topography-induced large amplitude interfacial waves \cite{bozoki} we have also demonstrated that linear three-layer theories, e.g. the one of Fructus \& Grue \cite{fg} may yield remarkably good fits to experimental data. Therefore, in the present work we focus on the applicability of linear theories to the dead water phenomenon in similarly stratified settings. Our aim here is to explore the $U$-dependence of the wavelength and amplitude of internal waves in the wake at fixed towing velocities and contrast the results with predictions of the linear theory for lee waves and for freely propagating three-layer interfacial waves. These measurements supplement the `constant force' experiments of Mercier et al. \cite{mercier} who have conducted state-of-the-art measurements with a presribed towing force (utilizing a falling weight) to propel a ship model in a rather similar set-up. 

The paper is organized as follows. Section 2 describes the experimental set-up and the applied data acquisition methods. Our results are presented in Section 3. The paper is then concluded with a brief discussion of the findings in Section 4.

\section{Experimental set-up and measurement methods}
\label{methods}

\begin{figure}[b!]
\noindent\includegraphics[width=\columnwidth]{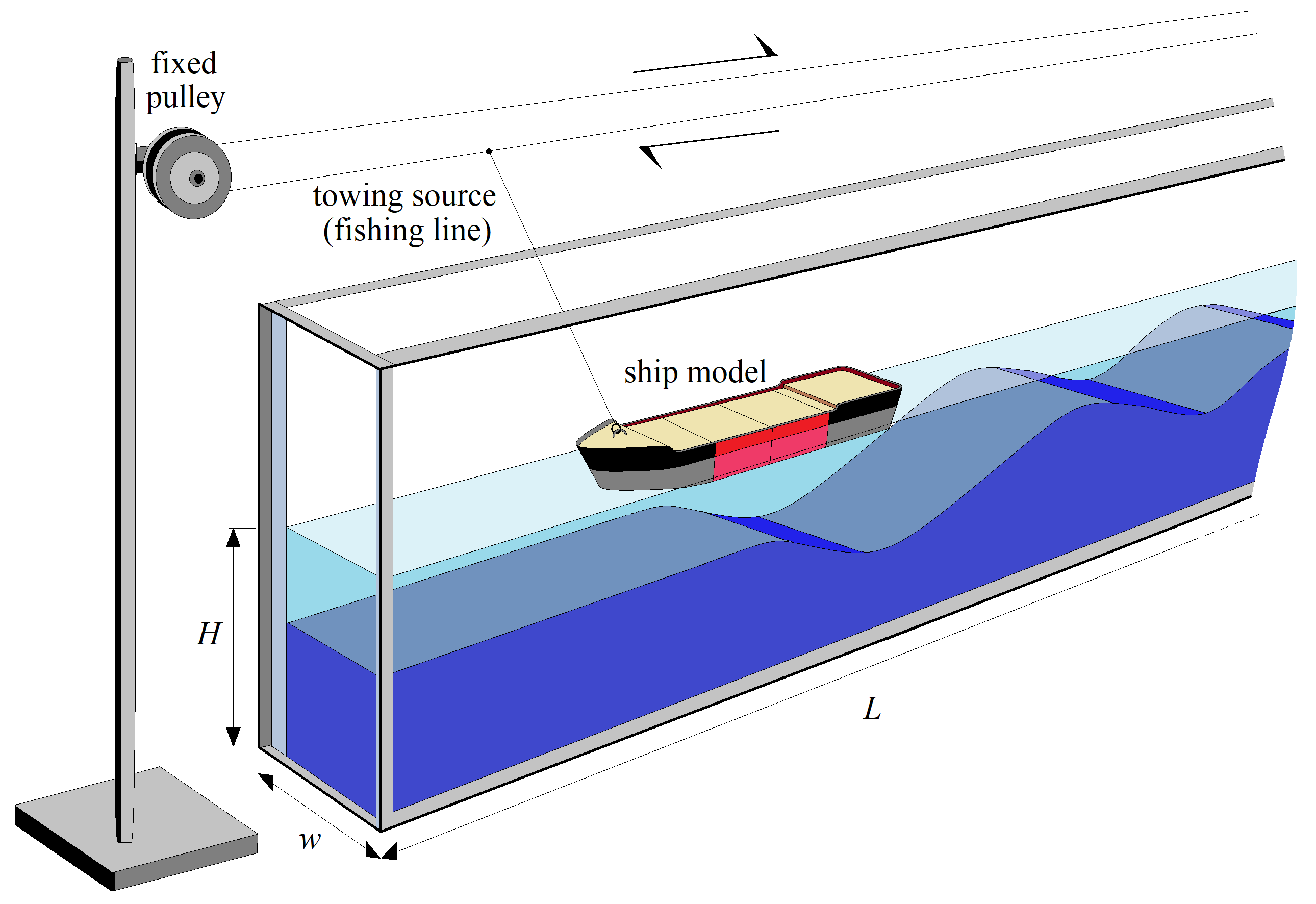}
\caption{The schematics of the set-up. The geometrical parameters of the tank are $L=239$ cm (total length, not fully shown), $w = 8.8$ cm, and $H = 12$ cm.}
\label{setup}
\end{figure}

\begin{figure*}[]
\centering
\noindent\includegraphics[width=0.7\textwidth]{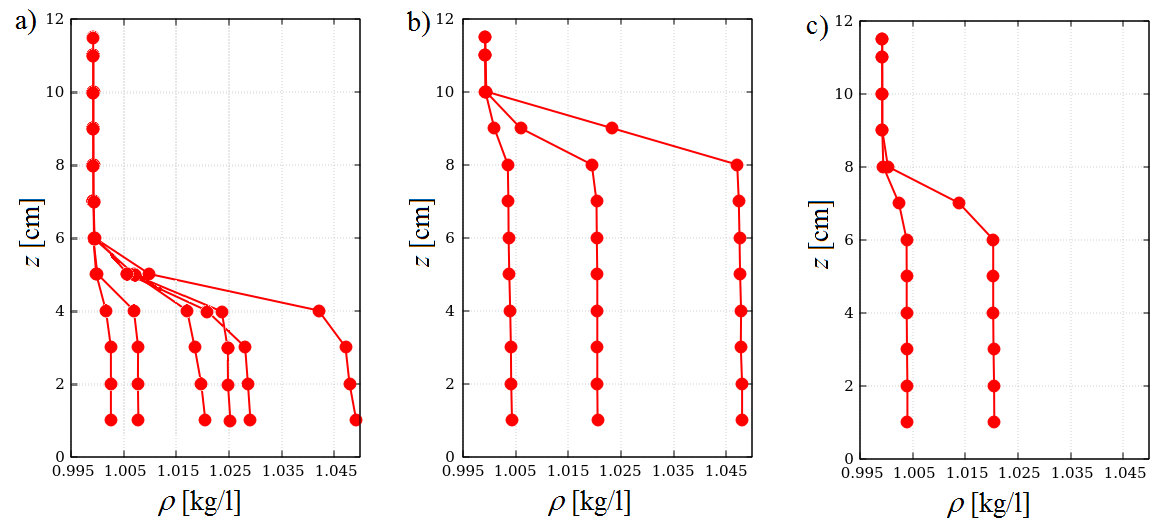}
\caption{Vertical density profiles of the experiments as measured a conductivity probe, with three layer thickness configurations and various maximum densities. (a) $h_1^{(2)} = 7$ cm, $h_2^{(2)} = 5$ cm; (b) $h_1^{(2)} = 3$ cm, $h_2^{(2)} = 3$ cm; $h_1^{(2)} = 5$ cm, $h_2^{(2)} = 7$ cm (cf. Table \ref{proftable}).}
\label{profiles}
\end{figure*}

The experiments reported here have been carried out in a rectangular laboratory tank made of transparent plexiglass. Its length and width are $L = 239$ cm $w = 8.8$ cm, respectively (see Fig. \ref{setup}). The tank was filled up to level $H = 12$ cm with density stratified water: the bottom domain contained saline water solution colored by red or blue food dye for the sake of visualization. Following the preparation of this layer, freshwater was poured through a sponge slowly onto the water surface in order to minimize mixing effects and to yield quasi-two layer density profiles, characterized by an approximately 2 cm thick region of steep density increase (see Fig. \ref{profiles}) or `gradient layer'. The temperature differences within the water body were negligible.

The properties of the 11 different stratification profiles are summarized in Table \ref{proftable}. 
$h_1^{(2)}$ and $h_2^{(2)}$ mark the effective thicknesses of the upper and bottom layers, respectively, using the two-layer approximation (hence the upper index `(2)'), i.e. assigning a jump-wise effective density change to the mid-height of the gradient zone. 
As a characteristic vertical scale in the system a `reduced thickness' $H_r$ can be introduced as the harmonic mean of $h_1^{(2)}$ and $h_2^{(2)}$, i.e. $H_r= (h_1^{(2)} h_2^{(2)})/ (h_1^{(2)}+ h_2^{(2)})$. Table \ref{proftable} also lists the values of the average density $\rho_2$ of the bottom layer and the corresponding two-layer linear interfacial wave velocity $c_0^{(2)}$ in the long-wave limit that reads as 
\begin{equation}
c_0^{(2)}=\sqrt{g \frac{\rho_2-\rho_1}{\rho_1} H_r},
\label{c0}   
\end{equation}
where $g$ denotes the gravitational acceleration and $\rho_1 \approx 1$ kg/l is the average density in the top (freshwater) layer \cite{sutherland}.

For a more precise treatment of the density profiles a three-layer approximation can also be applied \cite{fg}, in which the top, gradient and bottom layers (indexed with $j = 1,2,3$, respectively) are characterized by their approximate thicknesses $h_j^{(3)}$ and density gradients, or equivalently, their buoyancy (or, Brunt--V\"ais\"al\"a) frequencies $N_j$ that take the form
\begin{equation}
N_j\equiv \sqrt {-\frac{g}{\rho_0} \frac{d\rho}{dz}\Big|_j}.
\label{N}
\end{equation}
The values of $h_j^{(3)}$ and $N_j$ are obtained via piecewise linear regression to a given profile by determining the intersection points and slopes of the fitted lines. In the three-layer theory there is no such an explicit formula for the long wave velocities $c_0^{(3)}$ as in the two-layer approximation (\ref{c0}), as will be addressed later. Hence, the $c_0^{(3)}$-values in Table \ref{proftable} are numerical results.   

\begin{table*}[t]
\begin{center}
    \begin{tabular}{| c | c | c | c | c | c | c | c | c | c | c | c |} \hline
    Experiment series & $\#1$ &  $\#2$ &  $\#3$ &	 $\#4$ &	 $\#5$ &	 $\#6$ &  $\#7$ &	 $\#8$ &  $\#9$ &  $\#10$ &	 $\#11$ \\ \hhline{|=|=|=|=|=|=|=|=|=|=|=|=|}
      $h_1^{(2)}$ (cm) &	7 &	7 &	7 &	7 &	7 &	7 &	3 &	3 &	3 &	5 &	5\\ \hline
      $h_2^{(2)}$ (cm) &	5	&5&	5&	5&	5	&5	&9	&9&	9&	7&	7\\ \hline
      $H_r$ (cm) & 2.9 & 2.9 & 2.9 & 2.9 & 2.9 & 2.9 &	2.25 &	2.25 &	2.25 &	2.9 & 2.9\\ \hline
      $\rho_2$ (kg/l)&	1.019 & 1.029 & 1.008 & 1.003 & 1.047 & 1.025 & 1.004 & 1.020 & 1.048 & 1.004 & 1.020 \\ \hline
      $c_0^{(2)}$ (cm/s) & 7.30 & 8.92	&4.63 & 2.71 & 11.28 & 8.29 & 2.92 & 6.64 &	10.02 & 3.35 & 7.56 \\ \hhline{|=|=|=|=|=|=|=|=|=|=|=|=|}
      $h_1^{(3)}$ (cm) &	6 &	6 & 6 & 6 & 6 & 6 & 2 &	2 & 2 &	4 &	4\\ \hline
      $h_2^{(3)}$ (cm) &	2 & 3 &	3 & 3 &	3 &	2 &	2 & 2 &	2 &	2 &	2\\ \hline
      $h_3^{(3)}$ (cm) & 4 &	3 & 3 &	3 & 3 &	4 & 8 & 8 & 8 &	6 &	6\\ \hline
      $N_1$ (rad/s) & 0.18 & 0.32 & 0 & 0.03 & 0.16 & 0.23 &	0.10 & 0.27 & 0.37 & 0.24 & 0.55\\ \hline
      $N_2$ (rad/s) & 2.95 & 3.04 & 1.67 & 1.05 & 4.58 & 3.45 &	1.45 & 3.14	& 4.84 & 1.47	& 3.13\\ \hline
     $N_3$ (rad/s) & 1.05 & 0.69 & 0.15 & 0.0 &1.50 &	0.73 & 0.33 & 0.39 & 0.36 &	0.17 & 0.21\\ \hline
      $c_0^{(3)}$(cm/s) & 6.96 &	8.1	& 4.4 & 2.75 & 12.25 & 7.95	& 2.95 & 6.25 & 9.55 & 3.4 & 7.2\\ \hline
    \end{tabular}
\caption{Geometrical and physical parameters of the experiments for the two- and three-layer approximations (above and below the double line, respectively).}       
\label{proftable}
\end{center}
\end{table*}

\begin{table}[t]
\begin{center}
    \begin{tabular}{| c | c | c | c | c | c |}
    \hline
    configuration & $S1$ &  $S2$ &  $S3$ &	 $S4$ &	 $S5$  \\ \hline
    $d$ (cm) &	9.8& 16.2& 22.6& 29& 35.4 \\ \hline
    \end{tabular}
    \caption{The lengths of the used ship configurations.}
    \label{shiptable}
    \end{center}
    \end{table}

\begin{figure}[h]
\noindent\includegraphics[width=\columnwidth]{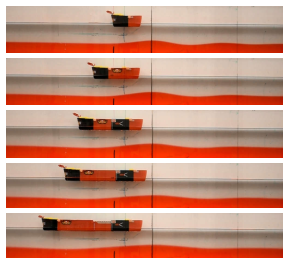}
\caption{The five different ship configurations of increasing length $d$, marked $S1$ to $S5$ (downward. cf. Table \ref{shiptable}). The bottom layer is visualized using red food dye. The snapshots are organized such that the first wave trough locations behind the ship models are underneath each other.}
\label{ships_fig}
\end{figure}

To capture dead water phenomenon we investigated the interfacial internal wave excitation behind towed LEGO\textsuperscript{TM} `tug boat' models (series 4005 and 4025) \cite{lego}. These toy ships have a modular design of a bow and a stern building block and four removable identical intermediate segments. Thus five configurations ($S_1$ to $S_5$) with different lengths $d$ could be investigated, as listed in Table \ref{shiptable} and shown in Fig. \ref{ships_fig}. The width of the models was 5.9 cm, comparable to tank width $w$. The ship model was towed by a (nylon) fishing line spanned horizontally above the water surface by ca. $10$ cm, as sketched in Fig. \ref{setup}, and driven by a DC motor whose voltage (and hence, the towing speed) could be adjusted between the experiment runs. The ship models' draught (i.e. the vertical distance between the waterline and the bottom of the hull) was found to be approximately $1$ cm in all cases. 

Each experiment was recorded with a HD video camera (at frame rate 50 fps and frame size $720\rm{px}\times 1280\rm{px}$) pointing perpendicularly to the sidewall close to the middle of the tank. To acquire the precise value of ship velocity $U$ and to obtain time series of vertical motion of the interface the video recordings were evaluated by \emph{Tracker}, an open source correlation-based feature tracking software \cite{tracker}. 

\section{Results}
\label{results}
\subsection{Qualitative description of the flow}
\label{res_i}
\begin{figure*}[h!]
\centering
\noindent\includegraphics[width=0.8\textwidth]{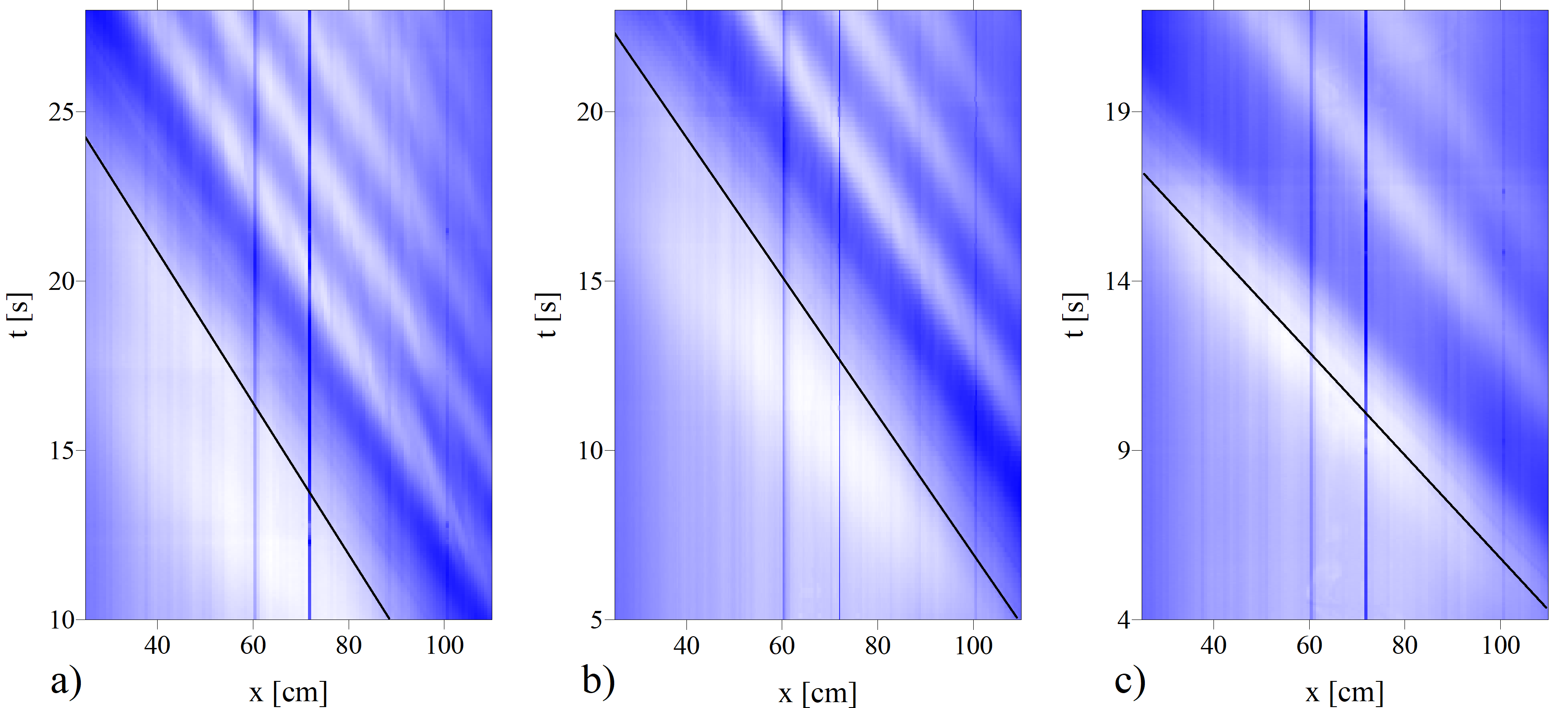}
\caption{Space-time plots of interfacial wave propagation behind the ship model. (Stratification profile \#11, ship configuration $S2$.)
The towing speeds are $U = 4.40$ cm/s (a) $U = 4.82$ cm/s (b) $U = 6.31$ cm/s (c) Horizontal position $x$ and time $t$ are measured from the left edge of the image and from the start of the ship motion, respectively. The coloring in each panel is normalized to the respective minimum and maximum values, thus amplitudes of the different cases cannot be compared to each other. The black lines represent the trajectories of the bow of the ship.}
\label{st}
\end{figure*}

As a ship model moves along the tank in the studied velocity range $U\in(1.3 ; 12.2)$ cm/s it generates pronounced waves on the internal interface, while the displacement of the free water surface remains negligible, as visible in Fig. \ref{ships_fig}. An important property of the observed dynamics is that the internal waves are following the ship and propagate at the same velocity as the ship itself. This is visualized in the space-time plots of Fig. \ref{st} for three different constant towing speeds $U$ (see caption).
In these diagrams the shading of a point at horizontal position $x$ and time $t$ is given by the sum darkness of the pixel column at $x$  as calculated from the video frame at time $t$ (e.g. the ones shown in in Fig. \ref{ships_fig}). Since the background of the tank is stationary throughout the videos, the spatial and temporal changes in darkness are attributed to internal waves. The trajectory of the bow of the ship is highlighted with solid black line in each panel.     

\begin{figure}[]
\noindent\includegraphics[width=\columnwidth]{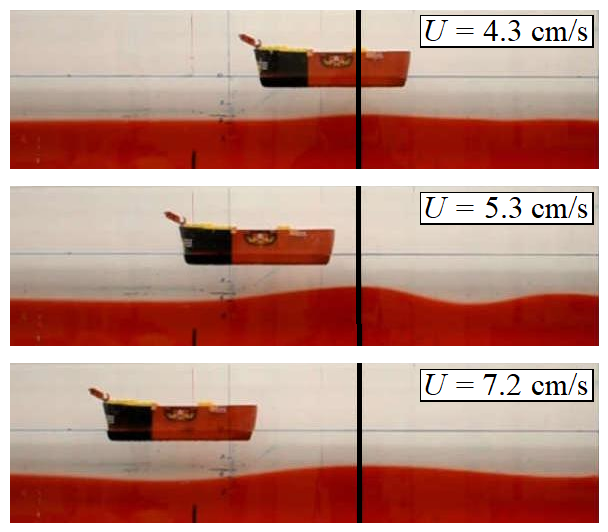}
\caption{Interfacial wave excitation behind ship model $S2$ for different towing velocities $U$ at stratification profile \#11. The snapshots are organized such that the first wave trough locations behind the ship models, marked by black vertical lines, are underneath each other.}
\label{qual_fig}
\end{figure}

The typical wavelength $\lambda$ and the characteristic amplitude $A$ of the internal waves are set by the towing speed $U$, the density profile $\rho(z)$, and the ship's length $d$. The $A(U)$ dependence is far from monotonous: for each stratification there exists an intermediate velocity $U^*$ at which internal waves of the largest amplitude develop. This `resonant' amplification is demonstrated with the video frames in Fig. \ref{qual_fig} for three values of towing velocity $U$, listed on the panels (stratification \#11, ship configuration $S2$). The snapshots are aligned such that the largest displacements of the density interface (marked with black vertical lines) are beneath each other for better comparability. The ship model's direction of motion is leftward in all images. 

At the smallest $U$ (uppermost panel) a wave crest appears at the interface beneath the ship model. For larger $U$'s a train of waves form (cf. Fig. \ref{st}) and the first crest shifts towards the stern of the ship, while the characteristic horizontal size (peak-to-peak wavelength $\lambda$) of the interfacial disturbance increases. Coincidentally, its vertical size (or, amplitude $A$) also increases and reaches a maximum at $U^*$, as captured in the second panel of Fig. \ref{qual_fig}. In the $U>U^*$ regime wavelength $\lambda(U)$ continues to increase, but amplitude $A(U)$ starts to decrease, as seen in the two bottom panels in Fig. \ref{qual_fig}.
In the following subsection we explore the parameter dependence of $U^*$ and the associated internal wave dynamics.        

\subsection{Parameter dependence of the resonance}
\label{res_ii}
\begin{figure*}[]
\centering
\noindent\includegraphics[width=0.7\textwidth]{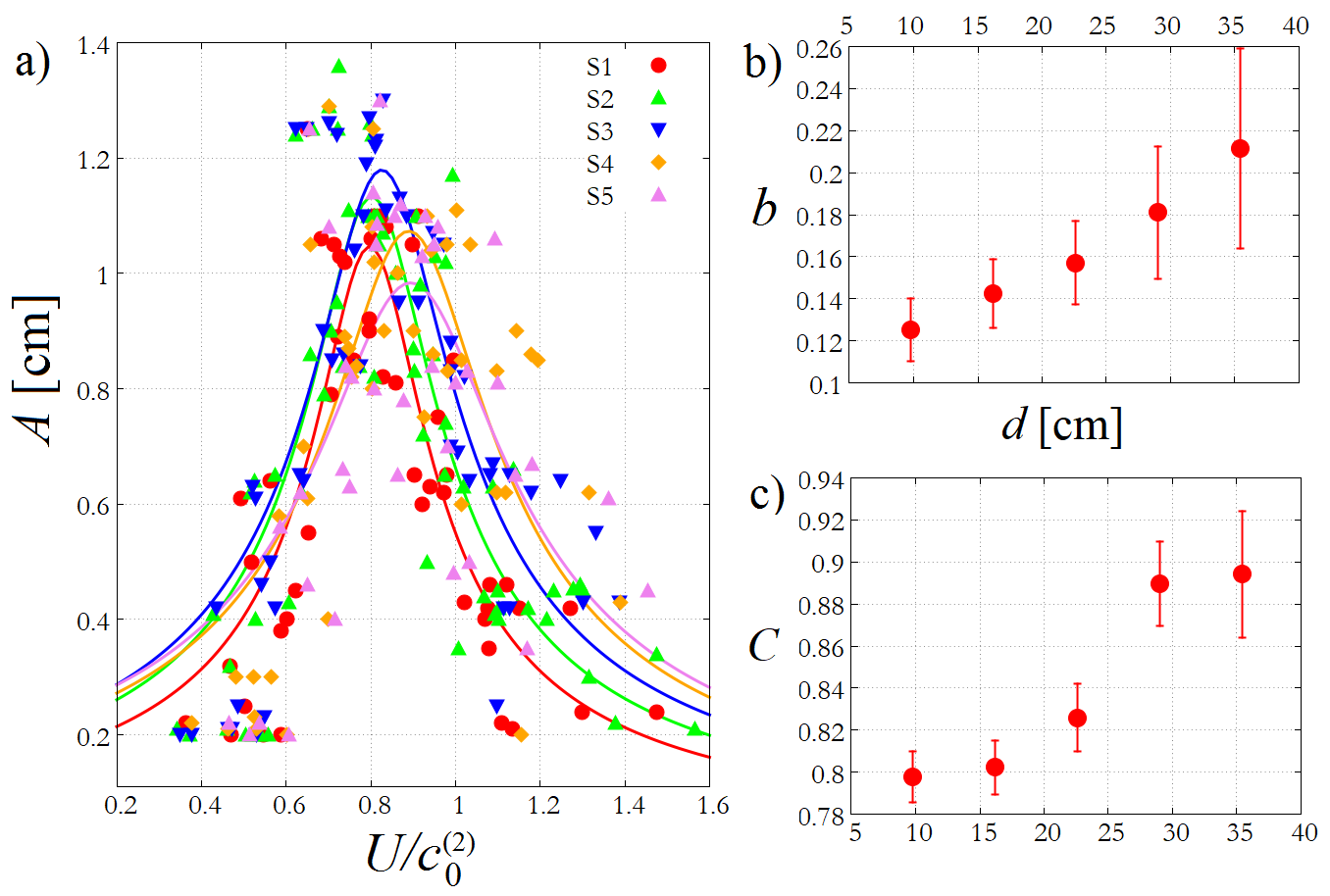}
\caption{(a) Maximum interfacial wave amplitudes $A$ as a function of nondimensional towing velocity $U/c_0^{(2)}$. The different symbols denote different ship configurations (see legend). The solid curves show the fits of the resonance function of equation (\ref{res_curve}) for each ship configuration, with the same color coding as the data points. (b) The width parameter $b$ of (\ref{res_curve}) as a function of ship length $d$. (b) The locations $C$ of the maxima of the fitted resonance curves (\ref{res_curve}) as a function of ship length $d$. The error bars represent fit errors in panels b and c.}
\label{lorentz}
\end{figure*}

The values of $U^*$ were found to range between $2.1$ cm/s and $9.1$ cm/s in the different experiments. Based on earlier results \cite{dw1,dw2} the relevant nondimensional velocity scale of the dead water problem is the internal Froude number i.e. the ratio of the ship speed $U$ and the two-layer long wave velocity $c_0^{(2)}$. Plotting the largest observed vertical interface displacement (peak amplitude) $A$ against $U/c_0^{(2)}$ for each experiment indeed yields a fairly good data collapse, as shown in Fig. \ref{lorentz}a with a peak at approximately 0.8. Each data point represents a single experiment run (towing) and the different symbols (and colors) indicate the various ship configurations used, as shown in the legend. 

Empirical `resonance curves' have been fitted to the above data for all five ship configurations separately, yielding the solid curves of Fig. \ref{lorentz}a (their coloring matches that of the corresponding data points).    
The function takes the 'usual' form of
\begin{equation}
A(U/c_0^{(2)}) = \frac{ab}{\sqrt{(U/c_0^{(2)}-C)^2+b^2}}.
\label{res_curve}
\end{equation}
Parameter $a$ measures the maximum of the curve, $b$ represents the `width' of the resonance and $C$ accounts for the position of the peak. The fitted values of $b$ and $C$ exhibit clear monotonous dependence on ship length $d$ as shown in Figs. \ref{lorentz}b and c. The $b(d)$ dependence even appears to be consistent with a linear scaling. The error bars represent the regression errors of the given parameters in both panels. 

It is to be noted that for all ship configurations the data points in Fig. \ref{lorentz}a manifestly scatter in a wider range around the fit in the $U > U^*$ (supercritical) range than for lower ship speeds. This implies that in this regime some physical parameter of the stratification profiles other than the Froude number $U/c_0^{(2)}$ becomes relevant, as will be addressed later.  

\subsection{Comparison with linear two- and three-layer theories}
\label{res_ii}
In what follows we compare the observed wave (and ship) speeds $U$ and wavenumbers $k$ to the available theoretical predictions of linear two and three layer theories.

\begin{figure}[]
\noindent\includegraphics[width=\columnwidth]{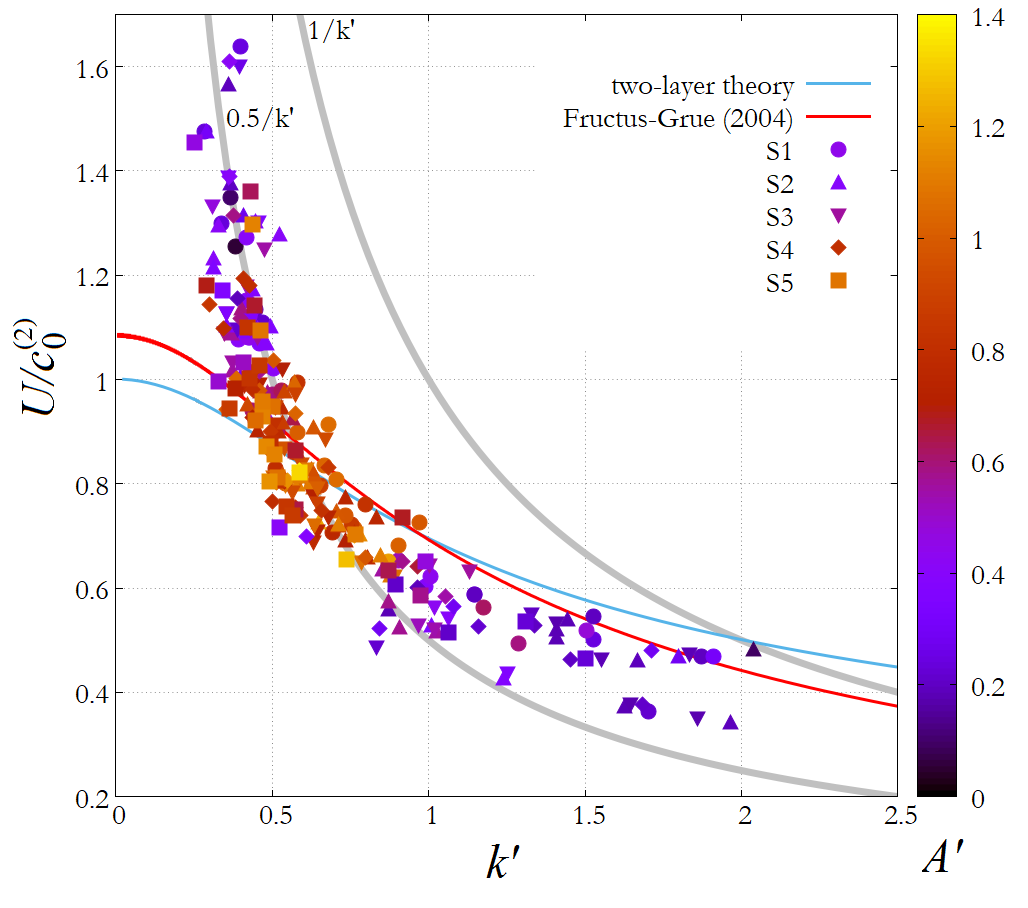}
\caption{(a)  Nondimensional towing (wave) speeds $U/c_0^{(2)}$ -- rescaled with the 2-layer long wave velocity, given by eq. (\ref{c0}) -- as a function of nondimensional wavenumber $k'=kH_r$. The symbol shapes mark different ship configurations (see legend), and the color scale marks non-dimensional amplitude $A'$, rescaled by the fitted maximum values of the corresponding resonance curves (cf. Fig. \ref{res_curve}). The blue curve represents the invariant (with respect to the nondimensional units used here) two dimensional velocity-wavenumber relation (\ref{linear_c_k}), the red curve denotes an exemplary three-layer relation, obtained for the parameters of experiment \#5, based on the implicit formula \ref{3layer}. The grey curves represent two hyperbolae,i.e. `iso-frequency curves' in the corresponding nondimensional time units, $y = 0.5/x$ and $y = 1/x$, respectively.
}
\label{ck}
\end{figure}

Assuming two homogeneous water layers of different densities separated by a sharp interface, the phase velocity reads as 
\begin{equation}
c^{(2)}(k)=\sqrt{\frac{g}{k}\frac{\rho_2-\rho_1}{\rho_1\coth(h^{(2)}_1k)+\rho_2\coth(h^{(2)}_2k)}},
\label{linear_c_k}
\end{equation}
where the notations are as introduced in Section \ref{methods} \cite{pedlosky}.
In the long-wave ($k \to 0$) limit the relationship takes the form of equation (\ref{c0}), therefore $c^{(2)}(0) \equiv c_0^{(2)}$. Expressing the velocities and wavenumbers in the problem's `natural' nondimensional units, i.e. $U/c_0^{(2)}$ and $kH_r$ (hereafter referred to as $k'$), respectively, maps equation (\ref{linear_c_k}) to the same invariant curve for all two-layer density profiles. This graph is shown with blue solid line in Fig. \ref{ck}, alongside the measured data points. The symbol shapes mark different ship configurations (see legend) and the coloring represents nondimensional wave amplitude $A'$, i.e. the maximum vertical displacement $A$ of the interface divided by the parameter $a$ of the fitted resonance curve of the given ship configuration (cf. Fig. \ref{lorentz}). $k'$ was calculated via measuring peak-to-peak wavelengths $\lambda = 2\pi/k$ between the second and third wave troughs. Note, that this method yields a considerable smaller value of $\lambda$ than the typical length of the first wave trough (cf. Fig. \ref{st}). Runs where no wave train developed were omitted from this analysis.

As expected, the linear two-layer theory systematically overestimates the wave speeds for larger values of $k'$. Interestingly, however, the best agreement between the data and the theory (in the $k' \approx 0.5$ domain) is observed at near-resonant wave speeds, i.e. where wave amplitudes $A'$ are large (see Fig. \ref{ck}).

For larger towing speeds ($U/c_0^{(2)}\gtrsim 0.8$) the observed wavenumbers remain larger than the predictions of the two-layer theory and exhibit a roughly inversely proportional scaling $U/c_0^{(2)}\propto k'^{-1}$ (see the gray hyperbolic guide curves in Fig. \ref{ck}) the implications of which will be discussed in subsection \ref{res_iii}.

The linear three-layer theory described in \cite{fg} has provided a quite accurate description of the observed interfacial velocities and wavenumbers in flow-topography interaction experiments \cite{bozoki}. This approximation assumes rigid top surface, small-amplitude waves, and three layers with depths $h_j^{(3)}$ ($j = 1,2,3$) and piece-wise linear stratification characterized by buoyancy frequencies $N_j$ (see eq. \ref{N} and Table \ref{proftable}). The dispersion relation $c^{(3)}(k)$ can then be derived numerically from the implicit equation
\begin{equation}
K_2^2-T_1 T_2 - T_1 T_3 - T_2 T_3 = 0,
\label{3layer}
\end{equation}
where $K_j = \sqrt{N_j^2/(c^{(3)})^2-k^2}$ is the vertical wave\-number in layer $j$ and $T_j = K_j \cot(K_j h^{(3)}_j)$.
The maximum wave speeds $c^{(3)}_0$ listed in Table \ref{proftable} corresponding to the long-wave ($k\to 0$) limit are also derived numerically from the above formulae.

To demonstrate the difference between the predictions of the two- and three-layer theories, the rescaled $c^{(3)}(k)$ curve calculated with the parameters of stratification profile \#5 (see Table \ref{proftable}) is added to Fig. \ref{ck} in the form of a red curve. Note, that unlike the aforementioned two-layer curve (blue) the three-layer one is not invariant at all in the units used here. For instance, in this exemplary case $c_0^{(3)}>c_0^{(2)}$ holds, but for many other profiles the sign would be reversed (cf. Table \ref{proftable}). 

\begin{figure*}[]
\noindent\includegraphics[width=\textwidth]{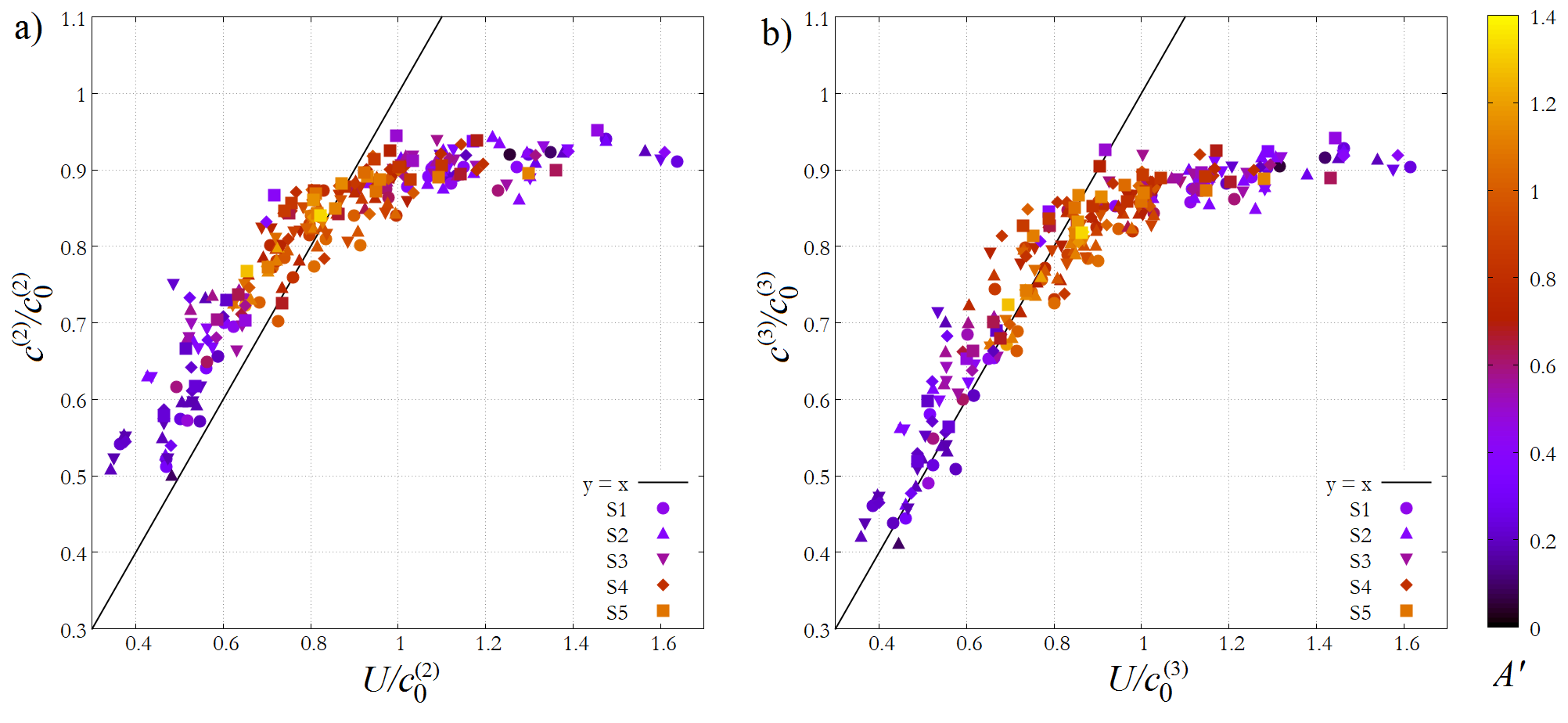}
\caption{Two-layer (a) and three-layer (b) theoretical velocities as a function of the measured towing (and wave-) speed $U$. The symbol shapes mark different ship configurations (see legend), and the color scale marks non-dimensional amplitude $A'$, rescaled by the fitted maximum values of the corresponding resonance curves (cf. Fig. \ref{res_curve}). The values on both the horizontal and vertical axes are rescaled with respect to the long wave limit velocities of the respective approximation (two layer for panel a, three-layer for panel b). The black solid curves represent the $y = x$ line.}
\label{corrplots}
\end{figure*}

In the correlation diagrams of Fig. \ref{corrplots}a and b we plot the theoretical phase velocities of the two- and three- layer models, respectively, against the measured wave speeds $U$ for each observed wavenumber $k$. In both panels the theoretical long wave velocity ($c_0^{(3)}$ or $c_0^{(2)}$) was used as the unit for nondimensionalization. Symbol shapes denote different ship configurations and the coloring represents rescaled amplitude $A'$ as in Fig. \ref{ck}. 

As noted before, the two-layer theory systematically overestimates the speeds in the $U/c_0^{(2)}\lesssim 0.8$ range (i.e. when $U\lesssim U^*$), thus the vast majority of the data points scatter above the $y = x$ line (black) in panel a. The three-layer theory, however, yields a fairly good match with the observations in the same subcritical regime. In the supercritical range, however, both two- and three-layer approximations break down entirely, as indicated by the deviation of the data points from the  $y = x$ line, implying that here another physical mechanism, besides the ship speed $U$, becomes relevant in the wavenumber selection.

\subsection{Lee-wave dynamics}
\label{res_iii}
A hyperbola in a velocity-wavenumber dispersion plot marks a constant frequency. For the nondimensional parameters of Fig. \ref{ck} the hyperbolic guide curves represent identical frequencies with respect to the stratification-dependent time unit $\sqrt{H_r\rho_1/(g(\rho_2-\rho_1))}$. Thus, the fact that the data points appear to follow hyperbolic scaling when $U > U^*$ implies that a characteristic frequency associated with the given density stratification $\rho(z)$ determines the observed wavenumbers.

\begin{figure}[]
\noindent\includegraphics[width=\columnwidth]{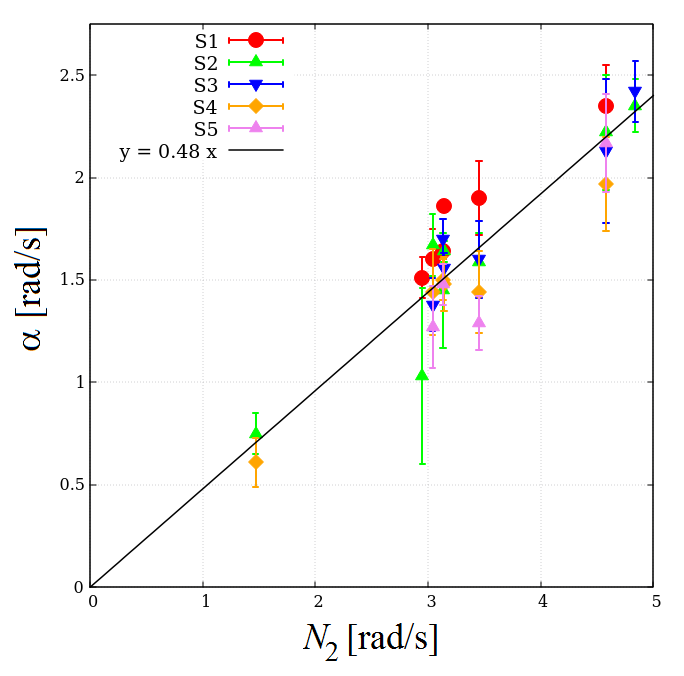}
\caption{Correlation plot between the intermediate layer buoyancy frequency $N_2$ and fitting parameter $\alpha$, obtained by regression to the $U(k)$ plots in the $U>U^*$ velocity range. The different symbols and colors represent different ship configurations (see legend). The error bars denote the fit errors (standard deviations) of $\alpha$. The slope of the black linear function is $0.48$.}
\label{N_vs_alpha}
\end{figure}

Such physically meaningful `eigenfrequencies' are the buoyancy frequencies $N_j$ of the layers (see Table \ref{proftable}), among which the mid-layer value $N_2$ is the largest in all cases. Fitting the function $\alpha/k$ to the dimensional $U(k)$ data in the $U>U^*$ (supercritical) range yields the empirical frequency parameter $\alpha$ that is plotted against $N_2$ in Fig. \ref{N_vs_alpha}. The error bars represent the regression errors and, as before, the different symbols mark various ship configurations. The scattering of the data points indicate a linear relationship $\alpha = 0.48(\pm 0.01)N_2$. The result of the fit is shown with a black solid line. 

The simplest (linear) theory that describes fixed frequency wave propagation behind an obstacle in a stratified fluid is that of \emph{lee waves} emerging at the downstream sides of mountain (or seamount) ridges \cite{sachs1,sachs2}. If the system is characterized by a single buoyancy frequency $N$ (linear stratification) and the obstacle is moving at velocity $U$ with respect to the medium then $U = N \cos(\phi)/k_{\rm lee}$ holds \cite{pedlosky}, where $\phi$ is the tilt of the wavenumber vector $\vec{k}_{\rm lee}$ from the horizontal ($|\vec{k}_{\rm lee}|= k_{\rm lee}$). Thus the maximum achievable $k_{\rm lee}$ becomes $k_{\rm lee}=N/U$, corresponding to fully horizontal wave propagation. 

As mentioned above, buoyancy frequencies $N_1$ and $N_3$ of the top and bottom layers are small in the studied density profiles, thus any lee wave activity is expected to be confined to the roughly 2 cm-thick intermediate layer which would then act as a `waveguide' \cite{bozoki} for the observed oscillation frequencies (i.e. above $N_1$ and $N_3$) at the interface. Since the layer thickness is an order of magnitude smaller than the typical wavelengths in the $U>U^*$ regime, $ \vec{k}_{\rm lee}$ is nearly horizontal here ($\phi\approx 0$). 

An important difference from the `traditional' lee wave setting is that in our configuration the ship model does not reach down to the intermediate layer, thus it is the depression of the interface underneath the ship and not the model itself that acts as a `moving obstacle' for the flow in the intermediate layer. The linear scaling presented in Fig. \ref{N_vs_alpha} appears to be consistent with the proposition that it is indeed lee wave-like dynamics that is observed in the supercritical regime, but -- due to the aforementioned peculiarities of the stratification profile -- not in its `classical' form. As an empirical correction we may therefore introduce an effective buoyancy frequency $N_{\rm eff}\approx 0.48 N_2$ to characterize the stratification, whose physical interpretation will be addressed in the Discussion. 

\begin{figure}[]
\noindent\includegraphics[width=\columnwidth]{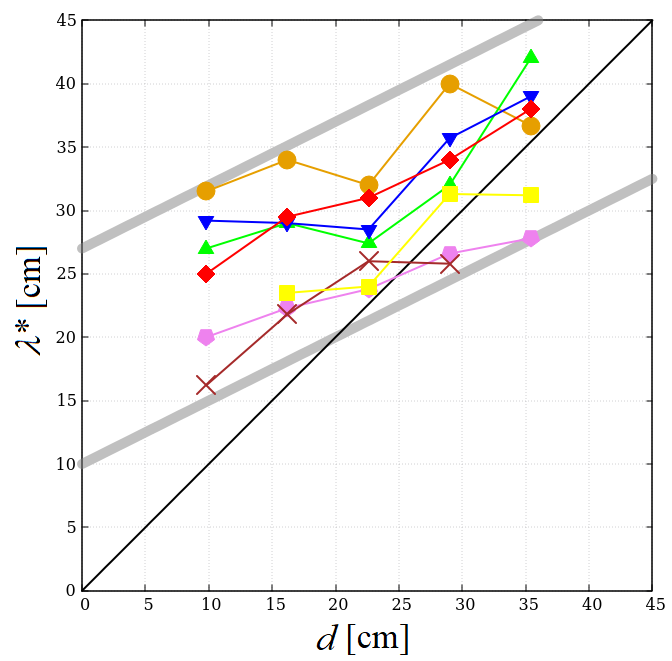}
\caption{Critical wavelength $\lambda^*$ (corresponding to the maximum amplitude) as a function of ship length $d$. The different symbold mark different stratifications. The slopes of the black and grey linear guidelines are $1$ and $0.5$, respectively.}
\label{lengthlambda}
\end{figure}

Finally, we investigated the critical wavelength $\lambda^* (\equiv 2\pi/k^*)$ corresponding to the maximum amplitude as a function of ship length $d$. The results are shown in Fig. \ref{lengthlambda} for seven different stratification profiles, marked by different colors and symbols. The solid black line marks $y = x$ and the gray guides are linear functions with a slope of 0.5 and a vertical offsets 10 cm and 27 cm. Apparently, the critical wavelength tends to increase with ship length, that roughly follows the empirical formula $\lambda^*\approx 0.5 d + f(\rho(z))$, implying that the ship configuration also plays a role in the wavenumber selection.

\section{Discussion and conclusions}
\label{disco}
Inspired by the historical work of Ekman analyzing the `dead water' phenomenon \cite{ekman} we have conducted laboratory experiments on wave excitation by a ship model towed at a fixed speed over a salinity stratified water body. 
We have analyzed the dependence of the interfacial wavenumber $k$ and peak-to-peak amplitude $A$ on the towing speed $U$, the stratification profile $\rho(z)$ and ship length $d$.

Due to the fact that in this setting the observed amplitudes are comparable to the characteristic vertical length scale $H_r$ of the problem, the excited waveforms can only be explained, if at all, by means of nonlinear wave theories (see, e.g. \cite{grue_sub,grue_super}). Yet, we deliberately focused our analysis on linear approximations in order to explore to what extent can these account for the basics of the observed dynamics, most notably the resonance-like amplitude amplification around $U/c_0^{(2)} = 0.8$ and the associated transition in terms of the $U(k)$ dependence. 

\begin{figure}[]
\centering
\noindent\includegraphics[width=0.75\columnwidth]{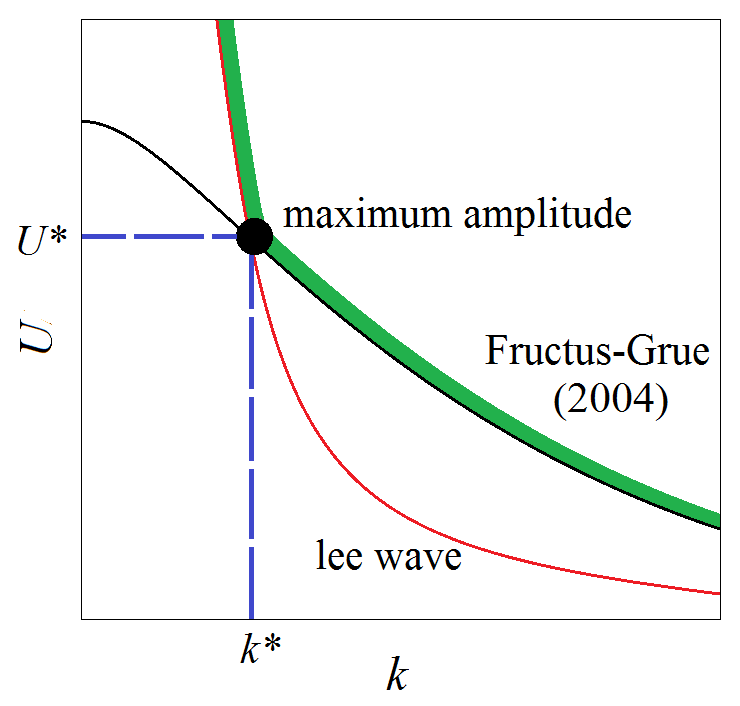}
\caption{The schematic wavenumber velocity diagrams of the lee waves (red curve), the 3-layer model of Fructus and Grue \cite{fg} (black), and the actually observable $U(k)$ domain (green) set by the maximum of the two competing $U(k)$ functions at each $k$. The intersection point $U^*(k^*)$  (black filled circle) fairly corresponds to maximum amplitude internal wave excitation.}
\label{discoplot}
\end{figure}

From the findings reported in the previous section it appears that the observable characteristic wavenumber $k$ at a given $U$ is set by the larger one of the corresponding  3-layer wavenumber $k^{(3)}(U)$ predicted by the linear approximation of Fructus and Grue \cite{fg} and the lee wavenumber $k_{\rm lee}(U)$ derived using `effective buoyancy frequency' $N_{\rm eff}\approx 0.48 N_2$. In other words, among the two competing mechanisms the one yielding shorter waves generates the first trough behind the ship model and, hence, sets the characteristic wavelength in the system. The crossing point of the two dispersion relations where $k_{\rm lee} = k^{(3)}$ holds is encountered around the critical towing speed $U^*$, as sketched in Fig. \ref{discoplot}. At this resonant wavenumber the two wave types would be superimposed onto each other resulting in an amplified interfacial wave excitation, that is confirmed by the observations. As a secondary effect, the selection of the resonant wavelength $\lambda^*$ was also found to be influenced by the length $d$ of the ship, as shown in Fig. \ref{lengthlambda} in unison with the increasing trend in the $U^*(d)$ dependence (Fig. \ref{res_curve}c).

The reason for the occurrence of the aforementioned factor of $0.48$ in the effective buoyancy frequency $N_{\rm eff}$ is unclear. Comparing this value to the internal wave dispersion relation $\omega=N_2 \cos(\phi)$ within the intermediate layer we find that it would imply a wave propagation whose lines of constant phase lie at an angle $\phi \approx 61 ^{\circ}$ to the vertical. However, we were not able to identify any geometrical constraint (e.g. one associated with the depression at the interface behind the ship) that would necessitate the presence of such a limiting angle. 

On the other hand, taking the total bottom-to-surface density difference $\Delta \rho$  and assuming linear stratification along the full depth $H$ gives a certain `mean buoyancy frequency' $N_{\rm m}=\sqrt{g/\rho_1\,(\Delta\rho/H)}$ which is found to follow the average scaling $N_{\rm m}\approx 0.46 (\pm 0.04) N_2$ for the profiles listed in Table \ref{proftable}, that is fairly close to $N_{\rm eff}$. Thus, the explanation may be that for long waves whose wavelength $\lambda$ exceeds $H$, the otherwise complex three-layer profile can be simply `averaged over' in terms of density gradients.

Our results clearly demonstrate the somewhat surprising and unexpected message that linear theories can occasionally be applied for the description of interfacial waves in such velocity and amplitude ranges that otherwise belong to the domain of nonlinear wave dynamics. It is clear, however, that linear models are insufficient to describe more complex features of the observed phenomena, e.g. waveforms, vorticity, etc. The authors also hope to increase awareness in the community about the dead water phenomenon, which -- despite being discovered over a century ago -- is still an interesting ground for theoretical, numerical, and experimental research.       

\begin{acknowledgements}
The essential contributions of Tam\'as T\'el and Bal\'azs Gy\"ure to the early, exploratory stage of dead water-related research at the K\'arm\'an Laboratory (which eventually led to the present paper) is highly acknowledged. We are also very grateful to Csilla Hrabovszki and her family who provided the `red' ship model. The work of K. M. is supported by the Stipendium Hungaricum Scholarship of the Tempus Public Foundation. This paper is also supported by the J\'anos Bolyai Research Scholarship of the Hungarian Academy of Sciences, the National Research, Development and Innovation Office (NKFIH) under Grant FK125024, 
and by the \'UNKP-18-4 New National Excellence Program (M.V.) of the Ministry of Human Capacities of Hungary.
\end{acknowledgements}


\begin{thebibliography}{}
\bibitem{glacier}
Parsmar, R., Stigebrandt, A. (1997). Observed damping of barotropic seiches through baroclinic wave drag in the Gullmar Fjord. Journal of Physical Oceanography, 27(6), 849-857.
\bibitem{ekman} Ekman, V. W. (1904). On dead water. Norwegian North Polar Expedition, 1893-1896, 1-150.
\bibitem{leewave_exp} Eiff, O. S., Bonneton, P. (2000). Lee-wave breaking over obstacles in stratified flow. Physics of Fluids, 12(5), 1073-1086.
\bibitem{knigge} Knigge, C., Etling, D., Paci, A., Eiff, O. (2010). Laboratory experiments on mountain‐induced rotors. Quarterly Journal of the Royal Meteorological Society: A journal of the atmospheric sciences, applied meteorology and physical oceanography, 136(647), 442-450.
\bibitem{vosper1999} Vosper, S. B., Castro, I. P., Snyder, W. H., Mobbs, S. D. (1999). Experimental studies of strongly stratified flow past three-dimensional orography. Journal of Fluid Mechanics, 390, 223-249.
\bibitem{vosper2004} Vosper, S. B. (2004). Inversion effects on mountain lee waves. Quarterly Journal of the Royal Meteorological Society, 130 (600), 1723-1748.
\bibitem{sachs1} Sachsperger, J., Serafin, S., Grubisic, V. (2015). Lee waves on the boundary-layer inversion and their dependence on free-atmospheric stability. Frontiers in Earth Science, 3, 70.
\bibitem{sachs2} Sachsperger, J., Serafin, S., Grubisic, V., Stiperski, I., Paci, A. (2017). The amplitude of lee waves on the boundary‐layer inversion. Quarterly Journal of the Royal Meteorological Society, 143(702), 27-36. 
\bibitem{yuan} Yuan, Y., Li, J., Cheng, Y. (2007). Validity ranges of interfacial wave theories in a two-layer fluid system. Acta Mechanica Sinica, 23(6), 597-607.
\bibitem{kdv} Apel JR (2003) A new analytical model for internal solitons in the 
ocean. J Phys Oceanogr 33(11):2247–2269
\bibitem{boschan} Boschan J, Vincze M, Janosi IM, Tel T (2012) Nonlinear resonance 
in barotropic–baroclinic transfer generated by bottom sills. Phys 
Fluids 24(4):046601
\bibitem{grue_sub} Grue, J. (2015). Nonlinear dead water resistance at subcritical speed. Physics of Fluids, 27(8), 082103.
\bibitem{grue_super} Grue, J., Bourgault, D., Galbraith, P. S. (2016). Supercritical dead water: effect of nonlinearity and comparison with observations. Journal of Fluid Mechanics, 803, 436-465.
\bibitem{grue_fram} Grue, J. (2018). Calculating FRAM's Dead Water. In The Ocean in Motion (pp. 41-53). Springer, Cham.
\bibitem{bozoki} Vincze, M., Bozoki, T. (2017). Experiments on barotropic–baroclinic conversion and the applicability of linear n-layer internal wave theories. Experiments in Fluids, 58(10), 136.
\bibitem{fg} Fructus, D., Grue, J. (2004). Fully nonlinear solitary waves in a layered stratified fluid. Journal of Fluid Mechanics, 505, 323-347.
\bibitem{mercier} M. J. Mercier, R. Vasseur, T. Dauxois (2011).
Nonlin. Processes Geophys., 18, 193-208.
\bibitem{sutherland} Sutherland BR (2010) Internal gravity waves. Cambridge University 
Press, Cambridge
\bibitem{lego} Instructions For LEGO 4005 Tug Boat (1982) available at:
\url{http://lego.brickinstructions.com/lego_instructions/set/4005/Tug_Boat_}
\bibitem{tracker} \url{http://physlets.org/tracker/}
\bibitem{dw1} Miloh, T., Tulin, M. P., Zilman, G. (1993). Dead-water effects of a ship moving in stratified seas. Journal of Offshore Mechanics and Arctic Engineering, 115(2), 105-110.
\bibitem{dw2} Motygin, O. V., Kuznetsov, N. G. (1997). The wave resistance of a two-dimensional body moving forward in a two-layer fluid. Journal of engineering mathematics, 32(1), 53-72.
\bibitem{pedlosky} Pedlosky J (2013) Waves in the ocean and atmosphere: introduction to 
wave dynamics. Springer, New York
\end{thebibliography}


\end{document}